# Voltage control of DC-DC Three level Boost converter using TS Fuzzy PI controller

Hajar Doubabi, Issam Salhi, Mohammed Chennani
Laboratory of Electric Systems and Telecommunications
Cadi Ayyad University
BP 549, Av Abdelkarim Elkhattabi
Marrakesh, Morocco
hajardoubabi@gmail.com, isalhi@yahoo.fr,
medchennani@gmail.com,

Najib Essounbouli
CReSTIC
Reims University
9 rue de Québec B.P 396, F-10026
Troyes cedex, France
najib.essounbouli@univ-reims.fr

*Abstract*— Appropriate control contributes essentially in the design of efficient DC-DC converters. With this intention, a study deals with the synthesis of a controller for DC-DC Three-level Boost converter (TLBC), has been addressed. The studied TLBC, known as nonlinear system, has been locally modeled using transfer function models. For instance, PI controllers were designed using the local models, and then they were combined using Takagi–Sugeno fuzzy (TSF) approach to form a TSF-PI controller. Simulation tests show the flexibility of the proposed controller, its rejection capability to different disturbances, and its ability to achieve the performance specification overall the wide operating range of the system.

*Keywords*— DC-DC Three-level Boost converter, Multi-model Identification, Fuzzy control, Takagi–Sugeno Fuzzy controller.

## I. Introduction

In the last three decades, Pulse Width Modulation (PWM) DC-DC converters have gained a very strong emphasis owing to their various features and their broad applicability. DC-DC converters are evolved to spread to almost every sector including transport, space and avionics, telecommunication, medicine and renewable energy. This is mainly thanks to new power semiconductor devices, new circuit structures and modern control techniques. Various DC-DC converters topologies have been presented in the literature and categorized according to their power conversion applications [1].

DC-DC Three-level Boost converter is a fundamental converter in power electronics that can efficiently steps up the input voltage. It is showing increasing popularity in power conversion applications due to its simplicity, high boost ability and flexibility. Fig. 1 shows the electrical schematic of a DC-DC TLBC. The DC-DC converters present a nonlinear dynamic behavior, which increases their modeling and control complexity [2].

In the literature, various studies have been conducted to investigate the DC-DC TLBC output voltage control. Different control methods have been proposed such as: (i) sliding mode control [3], (ii) model predictive control [4], and (iii) model-based control [5]. However, these approaches need complex mathematical analysis and require an accurate dynamic model to design the controller. Moreover, they may present major drawbacks such as the chattering problem in sliding mode control.

The classic control techniques, using PI or PID controller, are the most commonly used in the industry, as they are simple to synthetize, easy to implement, and could guarantee good performances in many cases. Nevertheless, for nonlinear system, these techniques have a limited validity (around local operating points) and cannot achieve good performances in the large domain of system operation.

Developing efficient control strategies for DC-DC converters based on fuzzy approach, has received an increasing interest in recent years [6-7]. This is owing to its ability of dealing with complex nonlinear system.

The so-called multi-model representation or local modeling is of great interest [8]. It allows the modeling of nonlinear systems based on linearization around different operating point. Further, the resulted models enable the use of classical control methods to design local controllers. The latters can be combined to fuzzy control method to design advanced control system.

In this paper, an effective and convenient feedback control strategy combining PI control and Takagi-Sugeno fuzzy (TSF) approach has been developed. It is mainly based on appropriately blending the local PI controllers together via the fuzzy membership functions in order to achieve a global controller (TSF-PI). The TSF-PI controller insures high control performance at all operating points in the wide range of the system operation [9]. This proposed control strategy has many merits including the possibility of using classic controllers (PI) to exploit their performances, simple handling of nonlinearity by using TSF approach, and no complicated mathematical computation is needed which enables simple implementation and low computational cost. This strategy has been examined in the literature and has shown improved performance in many processes, namely conventional DC-DC Boost converter [10] and micro hydro power plant [11].

In this paper, a TSF-PI controller, intended to control the output voltage of the TLBC, was developed. First, the linearization around specific operating points was performed. Second, the proper transfer functions, modeling the TLBC within the selected operating points, were set and validated. Accordingly, PI controllers were designed for each transfer function model. The resulted PI controllers were tested and have insured good performance, but only around the operating point for which they were designed. The obtained local PI controllers were blended together via TS fuzzy membership system to form the TSF-PI controller.

Simulation tests of the proposed controller were carried out in MATLAB/Simulink environment. The results reflect the TSF-PI controller advantages compared to traditional control in terms of robustness, flexibility and the ability to achieve specification over all range of operating points.

The paper is structured as follows. In Section II, the local TLBC identification based on transfer function models is addressed. Section III describes in detail the TSF-PI controller design and discusses the simulation results that validate the controller operation. Finally, a conclusion is given in Section IV.

## II. TLBC Modeling

As it can be seen from Fig. 1, the DC-DC TLBC consists of power switches $M_1$ and $M_2$, diodes $D_1$ and $D_2$, output voltage dividing capacitors $C_1$ and $C_2$, an inductor L, an equivalent series resistance (ESR) of inductor r, load resistor R, an input voltage $V_i$ and an output voltage $V_o$. For our study, the adopted TLBC parameters are given in table 1.

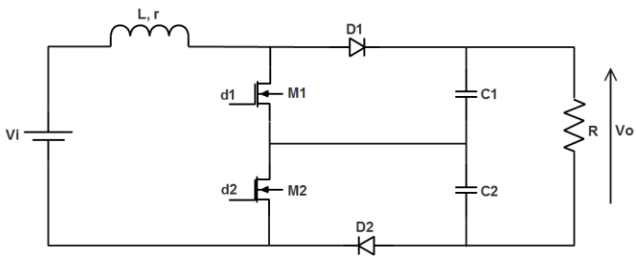

Fig. 1. Three-level Boost converter circuit

TABLE I. DC-DC TLBC PARAMETRS

| Parameter | Value | Unit |
|---|---|---|
| $V_i$ | 12 | V |
| L | 500 | μH |
| r | 8 | mΩ |
| $C_1$ | 100 | μF |
| $C_2$ | 100 | μF |
| R | 24.7 | Ω |
| $f_s$ | 32 | KHz |

$f_S$ is the switching frequency. $d_1$ and $d_2$ are defined as the duty ratios controlling the power switches $M_1$ and $M_2$ respectively. They lead to identical control signal but shifted by 180 degrees.

Based on the parameters listed in Table I, the TLBC circuit was implemented in MATLAB/Simulink as shown in Fig. 2. The corresponding operating characteristic (output voltage versus duty cycle) was plotted as illustrated in Fig. 3. This curve is a helpful tool, generally used to define the converter operation and to model its behavior, which leads to an effective converter identification. It can be inferred from this characteristic that beyond a maximum duty cycle, the converter would stop boosting. This is due to the resistance (ESR) in the inductor and power switches, and the diode voltage drop, that determine an upper limit on the duty cycle and thus the output voltage. Correspondingly, it is important to define this limit for every practical TLBC in order to prevent control loop instabilities [12].

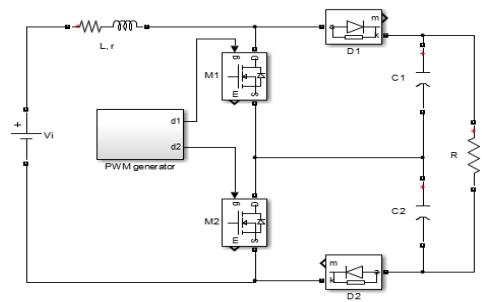

Fig. 2. TLBC circuit model in MATLAB/Simulink

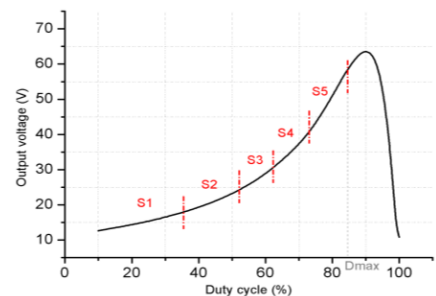

Fig. 3. Three-level Boost converter operating characteristic, divided into five linear subintervals

According to the depicted characteristic in Fig. 3, it is obvious that the TLBC presents a nonlinear behavior, which makes its modeling and control complicated. One of the simplest and most suitable approach that allows for a well-defined model enabling an accurate control is the small signal linearization. It can be achieved by identifying the system around specific operating points.

In our work, the TLBC operating characteristic is first divided into five linear subintervals [$S_1$, $S_2$, … , $S_5$] (as shown in fig. 3), and then transfer function models, corresponding to each subinterval, were extracted. The TLBC has two independent inputs, the input voltage (might be considered as a disturbance) and the duty cycle (that controls the converter). Hence, it is characterized by two transfer functions: the disturbance transfer function "$F_i$" and the process transfer function "$F_d$". The equivalent small signal block diagram of the TLBC in open loop, corresponding to each subinterval, is shown in Fig. 4.

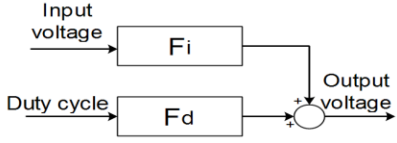

Fig. 4. Small-signal Block diagram of the open loop TLBC for a specific subinterval

The transfer function models of each of the five linear subintervals were identified based on the following procedure: (i) A small variation of the input voltage (of 1V) was applied in the open-loop TLBC circuit while the duty cycle was kept constant. The inputs and the output voltage were recorded; then (ii) A small variation of the duty cycle (of 1%) was applied in the open-loop TLBC circuit while the input voltage was kept constant. Those inputs variations engender sudden changes in the output voltage, as shown in Fig. 5 for the case of the sub-interval S1. The obtained data was then loaded and processed by MATLAB's System Identification Toolbox to deduce $F_d$ and $F_i$ corresponding to each linear subinterval. This toolbox enables a convenient identification of the system.

With regard to the transfer functions model, the number of poles generally depends on the number of dynamic elements (inductors and capacitors) that a converter comprises. Besides, the number of zeros can be changed to reach high 'fit to estimation data' and therefore reliable transfer functions. Specifically, the number of zeros was adopted based on tests that were performed using MATLAB's System Identification Toolbox. Different representations of both transfer functions $F_i$ and $F_d$, with no zeros to three zeros, have been examined. For instance, Table II presents the 'fit to estimation data' corresponding to each form of $F_i$ and $F_d$ for the case of the first subinterval $S_1$. As it was concluded, for all subintervals, the transfer function model with no zero has given the best fit to estimation data when representing to $F_i$, and the transfer function model with one zero has given the best fit to estimation data when representing to $F_d$. The resulted transfer function models are presented in Table III.

The transfer function models' responses have been compared to the ones of circuit model implemented in MATLAB/Simulink (model of the Fig. 2). The results confirmed that the transfer function model gives excellent approximation of the circuit model around the selected operating points.

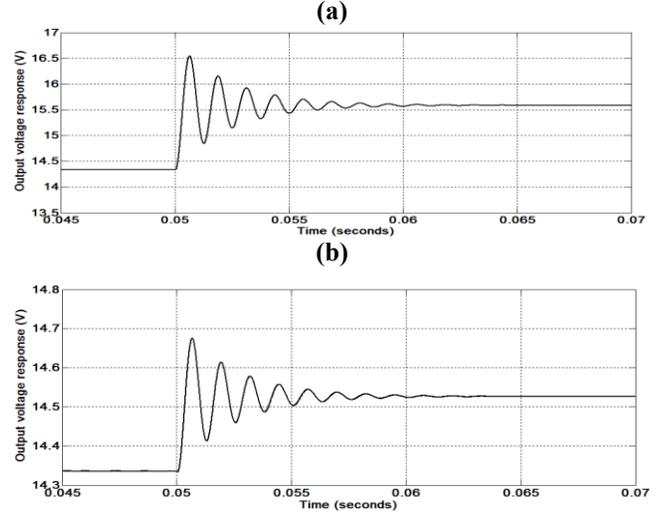

Fig. 5. Output voltage response to (a) input voltage variation (12V-13V) with duty cycle 20% (b) duty cycle change (20-21%) with an input voltage 12V.

TABLE II. THE 'FIT TO ESTIMATION DATA' (%) CORRESPONDING TO EACH FORM OF $F_{i1}$ AND $F_{d1}$

| $F_{i1}$ | | | | $F_{d1}$ | | | |
|---|---|---|---|---|---|---|---|
| no zeros | one zero | two zeros | three zeros | no zeros | one zero | two zeros | three zeros |
| **93.2** | < 0 | < 0 | 92.3 | 83.4 | **99.9** | 98.9 | 98.7 |

TABLE III. $F_i$ AND $F_d$ CORRESPONDING TO EACH SUB-INTERVAL $[S_1, S_2, ..., S_5]$

| | | |
|---|---|---|
| $S_1$ | $F_{i1}$ | $\dfrac{2.035 \cdot 10^{12}}{s^3 + 6.442 \cdot 10^4 s^2 + 7.87 \cdot 10^7 s + 1.629 \cdot 10^{12}}$ |
| | $F_{d1}$ | $\dfrac{-6.136 \cdot 10^8 s + 3.94 \cdot 10^{13}}{s^3 + 8.311 \cdot 10^4 s^2 + 5.087 \cdot 10^7 s + 2.067 \cdot 10^{12}}$ |
| $S_2$ | $F_{i2}$ | $\dfrac{1.541 \cdot 10^{12}}{s^3 + 6.509 \cdot 10^4 s^2 + 6.849 \cdot 10^7 s + 9.261 \cdot 10^{12}}$ |
| | $F_{d2}$ | $\dfrac{-8.927 \cdot 10^8 s + 3.928 \cdot 10^{13}}{s^3 + 8.302 \cdot 10^4 s^2 + 4.026 \cdot 10^7 s + 1.151 \cdot 10^{12}}$ |
| $S_3$ | $F_{i3}$ | $\dfrac{1.153 \cdot 10^{12}}{s^3 + 6.494 \cdot 10^4 s^2 + 6.223 \cdot 10^7 s + 5.205 \cdot 10^{11}}$ |
| | $F_{d3}$ | $\dfrac{-1.492 \cdot 10^9 s + 3.826 \cdot 10^{13}}{s^3 + 8.135 \cdot 10^4 s^2 + 3.376 \cdot 10^7 s + 6.276 \cdot 10^{11}}$ |
| $S_4$ | $F_{i4}$ | $\dfrac{7.729 \cdot 10^{11}}{s^3 + 6.535 \cdot 10^4 s^2 + 5.819 \cdot 10^7 s + 2.336 \cdot 10^{11}}$ |
| | $F_{d4}$ | $\dfrac{-3.313 \cdot 10^8 s + 3.65 \cdot 10^{13}}{s^3 + 8.024 \cdot 10^4 s^2 + 2.918 \cdot 10^7 s + 2.694 \cdot 10^{11}}$ |
| $S_5$ | $F_{i5}$ | $\dfrac{4.761 \cdot 10^{11}}{s^3 + 6.714 \cdot 10^4 s^2 + 5.74 \cdot 10^7 s + 8.761 \cdot 10^{10}}$ |
| | $F_{d5}$ | $\dfrac{-8.011 \cdot 10^9 s + 3.383 \cdot 10^{13}}{s^3 + 7.674 \cdot 10^4 s^2 + 2.588 \cdot 10^7 s + 8.904 \cdot 10^{10}}$ |

## III. PROPOSED CONTROL STRATEGY FOR THE TLBC

The main control objective is robustness regard to input variations and accurate reference tracking over the desired domain of operation. The control system synthesis generally builds on the model. Based on the converter model identification, the extracted transfer function models (TFMs) are used to design local controllers.

The controllers were designed around specific operating points. In other words, for each of the five TFMs, a PI controller is derived. Owing to its large applicability and simplicity of implementation, PI controller type is chosen. PI controller comprises an integral gain $K_i$ (to eliminate the output steady state error) and proportional gain $K_p$ (to reduce the rise time, which improve the system dynamic response). Fig. 6 shows the block diagram of the TLBC closed loop corresponding to each subinterval.

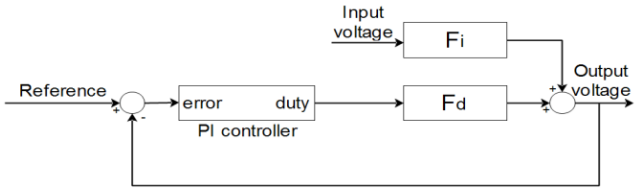

Fig. 6. Block diagram of the closed loop TLBC for a specific subinterval.

The controllers' parameters $K_p$ and $K_i$ have been set using 'pidTuner' application in MATLAB [2]. The controller in standard form presented by the equation (1) is selected. The parameters that can ensure a zero tracking error, low overshoot and fast settling time were defined. Table IV presents the chosen values of $K_p$ and $K_i$, corresponding to the PI controllers ($PI_1$, $PI_2$... $PI_5$) designed for the five TFMs, respectively.

$$C_{PI} = Kp(1 + Ki\frac{1}{s}) \qquad (1)$$

TABLE IV.  PI CONTROLLER PARAMETRS CORRESPONDING TO THE FIVE SUB-INTERVALS

|     | Kp | Ki |
| --- | --- | --- |
| $S_1$ | $5.24.10^{-6}$ | $1.42.10^6$ |
| $S_2$ | $2.93.10^{-6}$ | $1.72.10^6$ |
| $S_3$ | $1.64.10^{-6}$ | $1.21.10^6$ |
| $S_4$ | $7.34.10^{-7}$ | $1.38.10^6$ |
| $S_5$ | $2.60.10^{-7}$ | $1.18.10^6$ |

It should be mentioned that an anti-windup compensator in such feedback control system is inevitable. It is used to prevent the significant transient and overshoots ensued from the large growing of integrator state in PI controller. Extensive research has been dedicated to this subject [13].

Fig. 7 presents different scenarios demonstrating the reference tracking ability of the obtained PI controllers and their rejection capability of the input disturbance, but only when each PI controller is used in its specific sub-interval for which it was designed. The applied changes are: (i) an input variation occurs at 0.12s from 11V to 13V, (ii) a second input variation occurs at 0.25s from 13V to 11V and (iii) a reference voltage change at 0.36s. It is clearly seen that all PI controllers presents good performances and the system output tracks perfectly the reference voltage.

The controllers were also tested over other operating points. As an example, Fig.8 shows the output response of the closed loop system, with (a) $PI_1$ controller and (b) $PI_4$ controller, to changes in reference voltage following this sequence [15V-25V-42V-33V-50V] at 0.2s 0.4s 0.6s 0.8s respectively. An interesting aspect of this test is that the obtained results provided a global picture about the controllers' behavior within the whole operating region. As it can be seen, the use of the PI controllers out of their operating points presents slow control performances (a settling time of 110ms (see fig 8.(b))) and might lead the system to instability (see fig 8.(a)). The PI controllers have a limited validity domain and couldn't achieve the specification.

Accordingly, a Takagi–Sugeno fuzzy PI controller (TSF-PI), guaranteeing good performances overall operating range of the system, is developed. This controller is based on adjusting automatically the parameters ($K_p$ and $K_i$) of the above synthetized PI controllers depending on the working operating point. A diagram representing the closed loop system using the proposed controller is shown in Fig. 9.

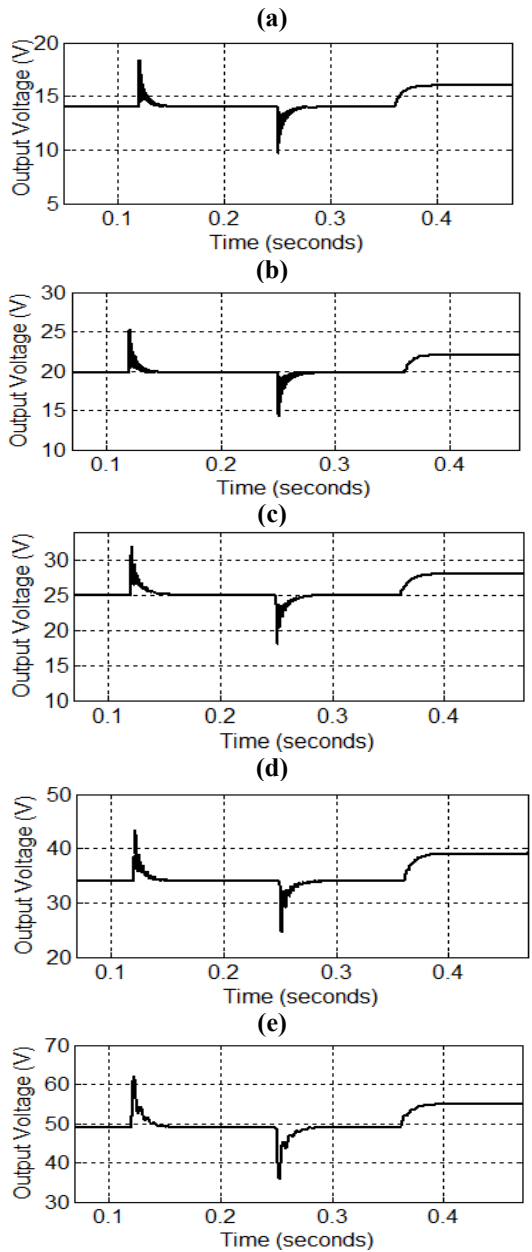

Fig. 7. Output voltage response to input voltage variation and reference voltage change for each PI controller: (a) PI$_1$ (b) PI$_2$ (c) PI$_3$ (d) PI$_4$ (e) PI$_5$.

For a given output voltage, the TSF-PI controller provides proper values of Kp and Ki. This is done by combining, in nonlinear form, the PI controllers' parameters given in Table III. The five linear subintervals [S1, S2… S5] were defined as the fuzzy sets, where S1= [12-18] S2= [18-24] S3= [24-31] S4= [31-40] S5= [40-57], and form the universe of discourse. The membership functions corresponding to these fuzzy sets are shown in Fig. 10. Based on Takagi–Sugeno Fuzzy method, the TSF-PI controller predicts accurate values of Kp and Ki that would allow suitable control.

Simulations testing the proposed controller are performed in MATLAB/Simulink environment. These simulations validate the utilized tuning strategy based on a TSF-PI controller in terms of robustness as well as reference tracking.

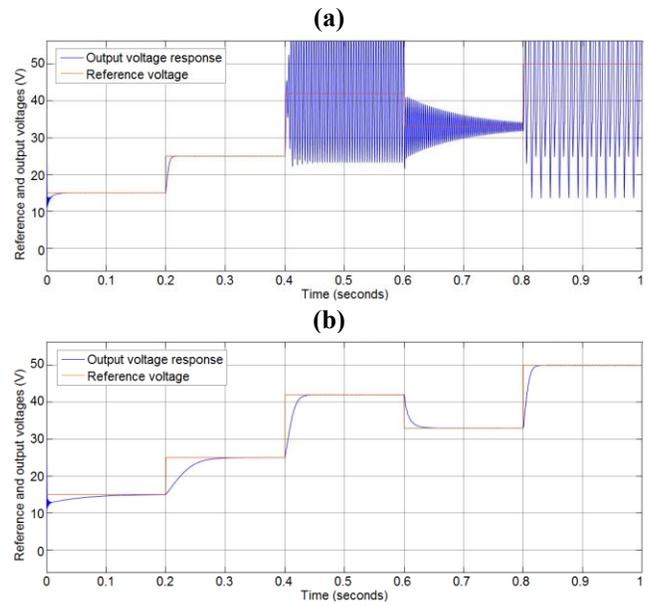

Fig. 8. The closed loop output response (a) with PI$_1$ controller (b) with PI$_4$ controller to reference voltage changes [15V-25V-42V-33V-50V].

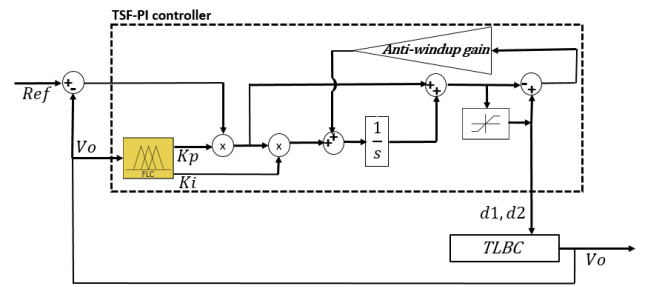

Fig. 9. Diagram of the closed loop system using TSF-PI controller

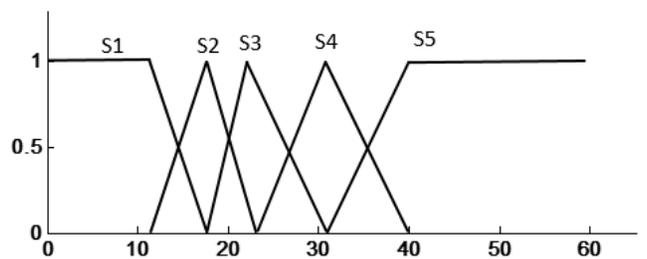

Fig. 10. Membership functions of the fuzzy sets of TSF-PI controller.

The first simulation test aims to study the effect of input voltage variation on the output voltage. Step changes of the input voltage were made where the input voltage followed this sequence [11V-12V-13V-12V-11V]. As shown in Fig. 11, the output voltage returned within milliseconds to its desired steady state value, which is in this case 17V. This test revealed that owing to the robustness of the designed controller, the output remains stable despite the input disturbances.

The second simulation test is related to the reference-tracking test overall the operating region of the converter. Reference has followed this sequence [15V-25V-42V-33V-50V] at 0.2s 0.4s 0.6s 0.8s respectively. It is clear from Fig. 12 that the proposed controller presents good performance tracking. The converter output tracks the reference (desired output) with good accuracy, no overshoots and the transient settling time is sufficiently small.

The obtained results show that the undesirable dynamic behavior of PI controllers out of their operating points illustrated in Fig. 8 is fixed by implementing the proposed TSF-PI controller. The TSF-PI controller remains valid overall operating points and enables undeniably appropriate control, which ensures the system reliability.

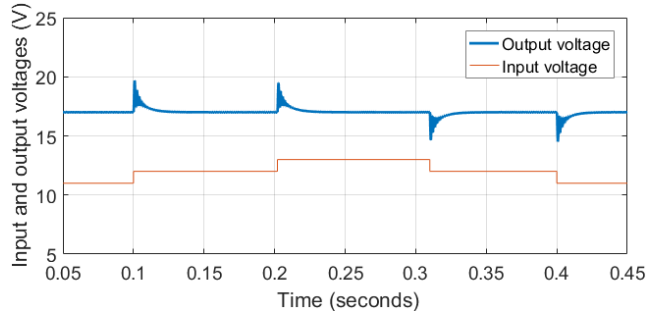

Fig. 11. Output voltage response in the presence of input voltage changes: [11V-12V-13V-12V-11V], where the reference voltage is 17V.

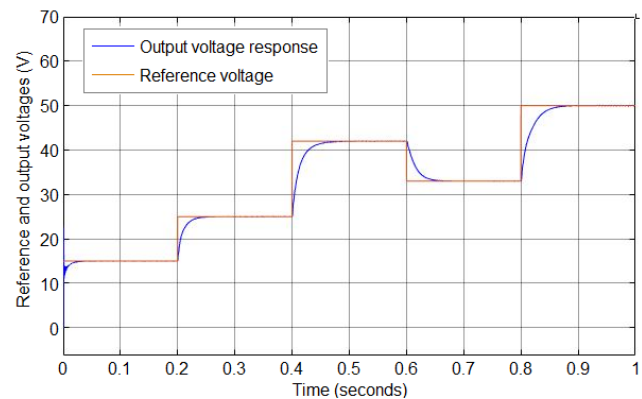

Fig. 12. Output voltage response for reference changes [15V-25V-42V-33V-50V], where the input voltage is 12V.

## IV. Conclusion

This paper deals with the design and implementation in MATLAB/Simulink of a Takagi–Sugeno fuzzy PI controller for a Three level Boost converter (DC-DC). A multi-model representation based on transfer function model was proposed to identify the system locally. Correspondingly, the system was first controlled using classical PI controllers that were designed for specific operating points. Simulation results have demonstrated that these controllers remain valid solely on the neighborhood of the operating point where they were designed for. Therefore, a TSF-PI controller with strong adaptability was synthetized. The simulations have emphasized good performance of the proposed controller under significant variations and disturbance over the entire domain of operation. Hence, a proper work of the TLBC is ensured.

## Acknowledgment

This work was supported by the National Center for Scientific and Technical Research (CNRST), Morocco. It is performed in the framework of VERES Project funded by the Research Institute in Solar Energy and New Energies (IRESEN), Morocco.